\documentclass[aps,reprint,apalike,assymb,twocolumn,10pt]{revtex4}
\usepackage{graphicx}       
\usepackage{amssymb}

\def\a{\alpha}
\def\b{\beta}

\def\g{\gamma}
\def\d{\delta}
\def\dd{\mbox{d}}

\def\s{\sigma}

\begin{document}

% You should use BibTeX and revtex.bst for references
%\bibliographystyle{apsrev}
\DeclareGraphicsExtensions{.pdf,.jpg,.tif} 
     
% Use the \preprint command to place your local institutional report
% number on the title page in preprint mode.
% Multiple \preprint commands are allowed.
%\preprint{}

%Title of paper
\title{Water alignment, dipolar interactions, and 
multiple proton occupancy during water-wire proton transport}

\author{Tom Chou\\
Dept. of Biomathematics and IPAM, Los Angeles, CA  90095-1766}

\date{\today}

\begin{abstract}

A discrete multistate kinetic model for water-wire proton transport is constructed
and analyzed using Monte-Carlo simulations. The model allows for each water
molecule to be in one of three states: oxygen lone pairs pointing leftward,
pointing rightward, or protonated (H$_{3}$O$^{+}$).  Specific rules for transitions
among these states are defined as protons hop across successive water oxygens. We
then extend the model to include water-channel interactions that preferentially
align the water dipoles, nearest-neighbor dipolar coupling interactions, and
coulombic repulsion. Extensive Monte-Carlo simulations were performed and the
observed qualitative physical behaviors discussed.  We find the parameters that
allow the model to exhibit superlinear and sublinear current-voltage relationships
and show why alignment fields, whether generated by interactions with the pore
interior or by membrane potentials {\it always} decrease the proton current. The
simulations also reveal a ``lubrication'' mechanism that suppresses water dipole
interactions when the channel is multiply occupied by protons. This effect can
account for an observed sublinear-to-superlinear transition in the current-voltage
relationship.  

\end{abstract}

% insert suggested PACS numbers in braces on next line
\vspace{2mm}

%\noindent Keywords: proton transport, 
%asymmetric exclusion process, water wire

\maketitle


\vspace{2mm}

\noindent {\bf INTRODUCTION}
\vspace{1mm}

The transport of protons in aqueous media and across membranes is a
fundamental process in chemical reactions, solvation, and pH regulation in
cellular environments \cite{MBOC,OSTER0}.  Proton transport in confined geometries is
also relevant for ATP synthesis \cite{BOYER} and light transduction by
bacteriorhodopsin \cite{LANYI}.  In this paper, we develop a lattice model for
describing proton transport in {\it one-dimensional}  environments. 
This study is motivated by numerous measurements of proton conduction
across channels embedded in lipid membranes
\cite{DEAMER91,BUSATH,BUSATH2,CUK97,DEAMER87,EISENMAN80}. 
Experiments are typically performed using membrane-spanning gramicidin
channels that  are only a few Angstroms in diameter.  This geometric constraint
imposes a single-file structure on the configurations of the interior water molecules
\cite{HIL78,HLA72}.  

\begin{figure}[h]
\begin{center}
\includegraphics[height=3.8in]{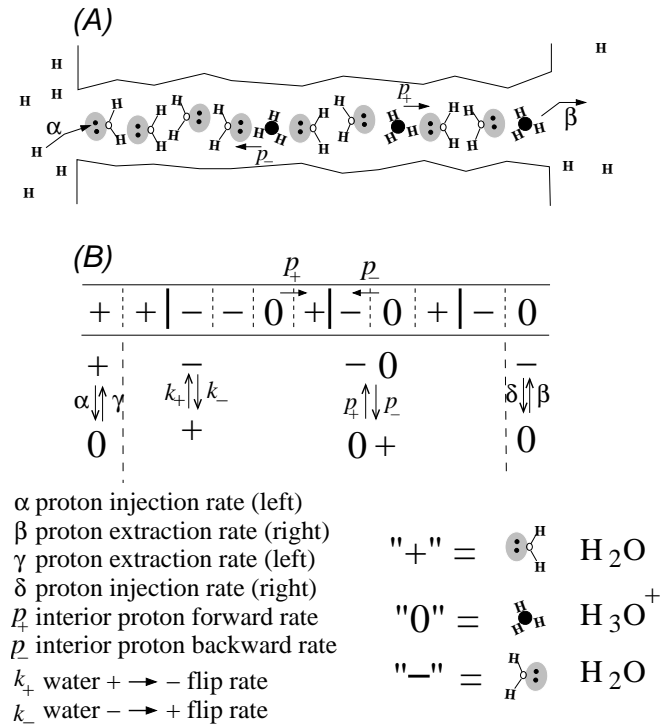}
\end{center}
\vspace{-2mm}
\caption{($A$) Schematic of an $N=11$, three-species exclusion model 
that captures the steps in a Grotthuss mechanism of proton transport
along a water wire. For typical ion channels that span 
lipid membranes, $N\sim 10-20$. The transition rates are labeled in (B) and in the 
legend. Water dipole kinks are denoted by thick lines.}
\label{fig1}
\end{figure}

Under the same electrochemical potential gradients, 
conduction of protons across ion channels occurs at a rate typically an order of
magnitude higher than that of other small ions. This supports a ``water-wire'' mechanism
\cite{DEAMER91,Nag78,Nag83,Nag87}, first proposed by Grotthuss
\cite{AGMON,GROTTHUSS}. Across a water-wire, protons are shuttled across lone
pairs of water oxygens as they successively protonate the waters along the single-file
chain.  However, since the hydrogens are indistinguishable, any one of the hydrogens in a
water cluster ({\it e.g.}, any of the three hydrogens on a hydronium) can hop forward along
the chain to protonate the next water molecule or cluster of water molecules (cf. Fig.  \ref{fig1}). This
mechanism naturally allows much faster overall conduction of protons compared to other
small ions which have to wait for the entire chain of water molecules ahead of it to fluctuate across the
pore in order to traverse the channel.

A peculiar feature of measured current-voltage relationships is a crossover from sublinear to
superlinear behavior as the pH of the reservoirs is lowered.  Measurements by Eisenman
\cite{EISENMAN80} were carried out in symmetric solutions in the 1-3 pH range, and the
results were recently reproduced by Busath {\it et al.} \cite{BUSATH} and Rokitskaya {\it et
al.} \cite{ROKIT}.  These experiments were performed using simple, relatively featureless
gramicidin A (gA) channels. One leading hypothesis is that the nonlinear proton
current-voltage relationships arise from the intrinsic proton dynamics within 
such simple channels.  Specifically, multiple proton occupancy and repulsion among
protons within the channel may give rise to the observed nonlinearity
\cite{HIL78,PHILLIPS,SHU}.

There have  been a number of recent theoretical studies of water-wire proton conduction. Extensive
simulations on the quantum dynamics of proton exchange in essentially small, representative water
clusters in vacuum have been used to predict microscopic hopping rates between water clusters
\cite{QUANTUM3,CHENG,PARRINELLO,Berendsen95,Klein98,Klein96,SCHMITT}.  Pom\`{e}s and
Roux \cite{POM96} have performed classical molecular dynamics (MD) simulations on water-channel
interactions, proton hopping, and water reorientation.  They derive effective potentials of mean force
describing the energy barriers encountered by a single proton within the pore.  Since MD simulations
are presently limited to only processes that occur over a few nanoseconds, none of these
computational methods are efficient at probing very long time, steady-state transport behavior.  On a
more macroscopic, phenomenological level, Sagnella {\it et al.} \cite{Voth96} and Schumaker {\it et al.}
\cite{Schumaker99,SHU} have considered a the long-time behavior of a single proton and dipole
"defect" diffusing in a single-file channel.  The parameters used in these studies, including effective
energy profiles and kinetic rates, were derived from MD simulations.  Although the basic underlying
structure assumed by all of these transport models qualitatively resembles the Grotthuss mechanism,
they have not addressed multiple proton occupancy.

In this paper, we will explore the intrinsically nonlinear proton dynamics along a single-file
water-wire.  We use a dynamical lattice model that defines 
the discrete structural states of the water-wire  that approximate 
the continuous molecular orientations.  Although the lattice model provides a 
different approach from MD simulations, it is more amenable 
to analysis at longer time scales, yet is connected to the microphysics 
inherent in MD simulations provided a consistent correspondence
between  the parameters is made. Rather than enumerating  all
possible molecular configurations, our lattice approach is resembles that 
developed for molecular motors \cite{FISHER}, mRNA translation \cite{MRNA1,MRNA2},
traffic flow \cite{TRAFFIC1,TRAFFIC2}, and ion and water transport in single-file channels
\cite{CHOU98,CHOU99,CHOULOHSE99}. Here, the proton
occupancy along the water-wire will be self-consistently determined by the prescribed
lattice dynamics.  The parameters used in our model are transition rates among discrete
states that in principle can be independently computed from relatively short-time MD
simulations. Despite the approximations inherent in our discrete model, 
it qualitatively treats the effects of proton-proton repulsion and water-water
dipole interactions.

%In the next section, we formulate the lattice model and the associated
%dynamical rules.   Microscopic, physical assumptions that need
%to be invoked are also presented.  Both a three site model and the
%implementation of a Monte-Carlo simulation also discussed. The three site
%model is solved numerically to gain qualitative insight and serve as a check for
%the MC simulations of longer water wires. Monte-Carlo simulation results are
%presented in the Results and Discussion.  The effects of the various nearest
%neighbor interactions are systematically studied in a series of plots of the
%simulation  results.

\vspace{4mm}
\noindent {\bf MODEL AND METHODS}
\vspace{1mm}

Qualitatively, protons hop from oxygen to oxygen during transport. The successive hops
clearly do not have to involve an individually tagged proton; in this respect, proton currents
resemble electrical conduction in a conductor.  Many measurements of proton conduction
across membranes are performed on the gramicidin model system.  The interior diameter of
gramicidin A (gA) is $\sim 3-4$ Angstroms and can only accommodate water in a single file
chain.  Although the number of water molecules in this chain is a fluctuating quantity, their
dynamics in and out of the channel will be assumed to be much slower than that of their
orientational rearrangements and proton hopping \cite{HUMMER01,HUMMER03}. We can
thus treat the water wire as containing a fixed, average number of water molecules. Within
typical transmembrane channels, are $N \approx 8-26$ single-file waters
\cite{LEVITT78,WU}.

Figure 1$A$ shows a schematic of our model. We first assume that each ``site'' along the
pore is occupied by a single oxygen atom which may either be part of neutral water,
H$_{2}$O, or a hydronium (H$_{3}$O$^{+}$) ion.  Although protonated oxygens in bulk are
often associated with larger complexes such as H$_{5}$O$_{2}^{+}$ (Zundel cation), or
H$_{9}$O$_{4}^{+}$ (Eigen cation), in confined geometries, the formation of the 
larger complexes is suppressed \cite{LBELL}.  Furthermore, we will show  that our discrete
model depicted in Fig. 1$A$ can incorporate the dynamics of reactive proton transfer among
transient clusters by an appropriate redefinition of a lattice site to contain the entire 
cluster.

Neutral waters have permanent dipole moments and electron lone-pair
orientations that can rotate thermally. For simplicity, we bin all water dipoles (hydrogens)
that point towards the right  as ``$+$'' particles, while those pointing more or less to the left
are denoted ``$-$'' particles.  The singly protonated species H$_{3}$O$^{+}$ is
hybridized to a nearly planar molecule.  Therefore, we will assume that hydronium ions are
symmetric with respect to transferring a proton forward or backwards, provided the adjacent
waters are in the proper orientation and there are no external driving forces (electric fields). 
Each lattice site can exist in only one of three states: $0, +$, or $-$, corresponding to
protonated, right, or left states, respectively. Labeling the occupancy configurations $\s_{i}
= \{-1, 0, +1\}$, allows for fast integer computation in simulations.  

In addition to  proton exclusion, the transition rules are constrained by the orientation of the waters at
each site and are defined in Fig. 1$B$.  A proton can enter the first site ($i=1$) from the left reservoir
and protonate the first water molecule with rate $\a$ only if the hydrogens of the first water are
pointing to the right (such that its lone-pair electrons are left-pointing, ready to accept a proton). 
Conversely, if a proton exits from the first site back into the left reservoir (with rate $\gamma$), it
leaves the remaining hydrogens right-pointing.  In the pore interiors, a proton at site $i$ can hop to
the right(left) with rate $p_{+}(p_{-})$ only if the adjacent particle is a right(left)-pointing,
unprotonated water molecule. If such a transition is made, the water molecule left at site $i$ will be
left(right) pointing.  Physically, as a proton moves to the right, it leaves a wake of $-$ particles to its
left.  A left moving proton leaves a trail of $+$ particles to its right.  These trails of $-$ or $+$ particles
are unable to accept another proton from the same direction.  Protons can follow each other
successively only if water molecules can reorient such that these trails of $+$'s or $-$'s are thermally
washed out.  Water reorientation rates are denoted $k_{\pm}$ (cf. Figs.  1$B$ and 2). Protons at the
rightmost end of the water wire (at site $i=N$), exit with rate $\beta$, which is different from $p_{+}$
since the local microenvironment ({\it e.g.}, typical distance to acceptor electrons) of the bulk waters
that accept this last proton is different from that in the pore interior.  From the right reservoir, protons
can hop back into the water wire with rate $\delta$ if a water in the ``$-$'' configuration is at site 
$i=N$.  The entrance rates $\alpha$ and $\delta$ are functions of at least  the proton concentration in
the respective reservoirs.  Figure 2 shows a representative time series of the evolution of a specific
configuration.  The rate-limiting steps in steady-state proton transfer across biological water channels
are thought to be associated with water flipping \cite{POM98}.  

\begin{figure}
\begin{center}
\includegraphics[height=2.2in]{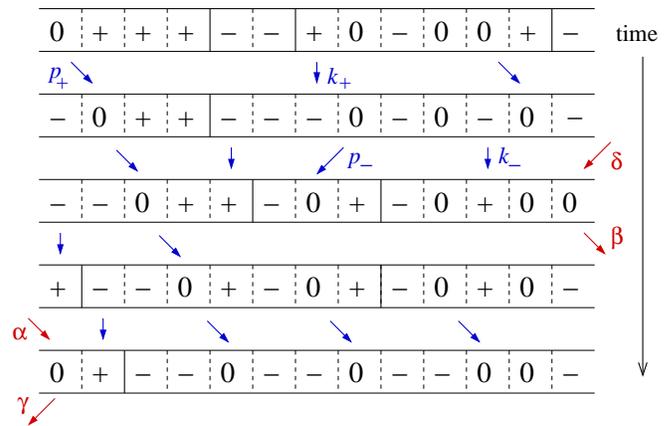}
\end{center}
\vspace{-2mm}
\caption{A time series depicting 
a number of representative transitions obeying the 
dynamical constraints of our model. A proton (0) at site $i$ can
move to the right with rate $p_{+}$ only if site $i+1$ is occupied by a properly 
aligned (lone-pair electrons pointing to the
left)  water molecule (+).  When a proton leaves site $i$ to the 
right, it leaves behind a water in state ``$-$'', 
with lone pair electrons pointing to the right. Protons at site $i$ can also 
move to the left with rate $p_{-}$ if site $i-1$ is a water in the ``$-$'' state. 
In this case, a water is left behind at behind site $i$ in the ``$+$'' state. The neutral water 
molecules must flip ($+ \leftrightarrow -$) in order for a nonzero 
steady-state current to exist.}
\label{series}
\end{figure}

The lattice discretization for individual H$_{3}$O$^{+}$
ions need not be interpreted literally. Larger complexes can be effectively modeled 
by reinterpreting $p_{\pm}, k_{\pm}$, and the basic unit of hopping for the proton.
For example, if certain conditions obtain, where ions are predominantly 
two-oxygen clusters (H$_{5}$O$_{2}^{+}$), we defined each pair of waters as occupying a 
single lattice site, $k_{\pm}$ as an effective reorientation time for the 
following pair of waters, and $p_{\pm}$ as the hopping rate to an adjacent 
oxygen lone-pair. The Grotthuss water-wire mechanism is qualitatively preserved as 
long as the proper identification with the microphysics is made.

\begin{figure}
\begin{center}
\includegraphics[height=3.0in]{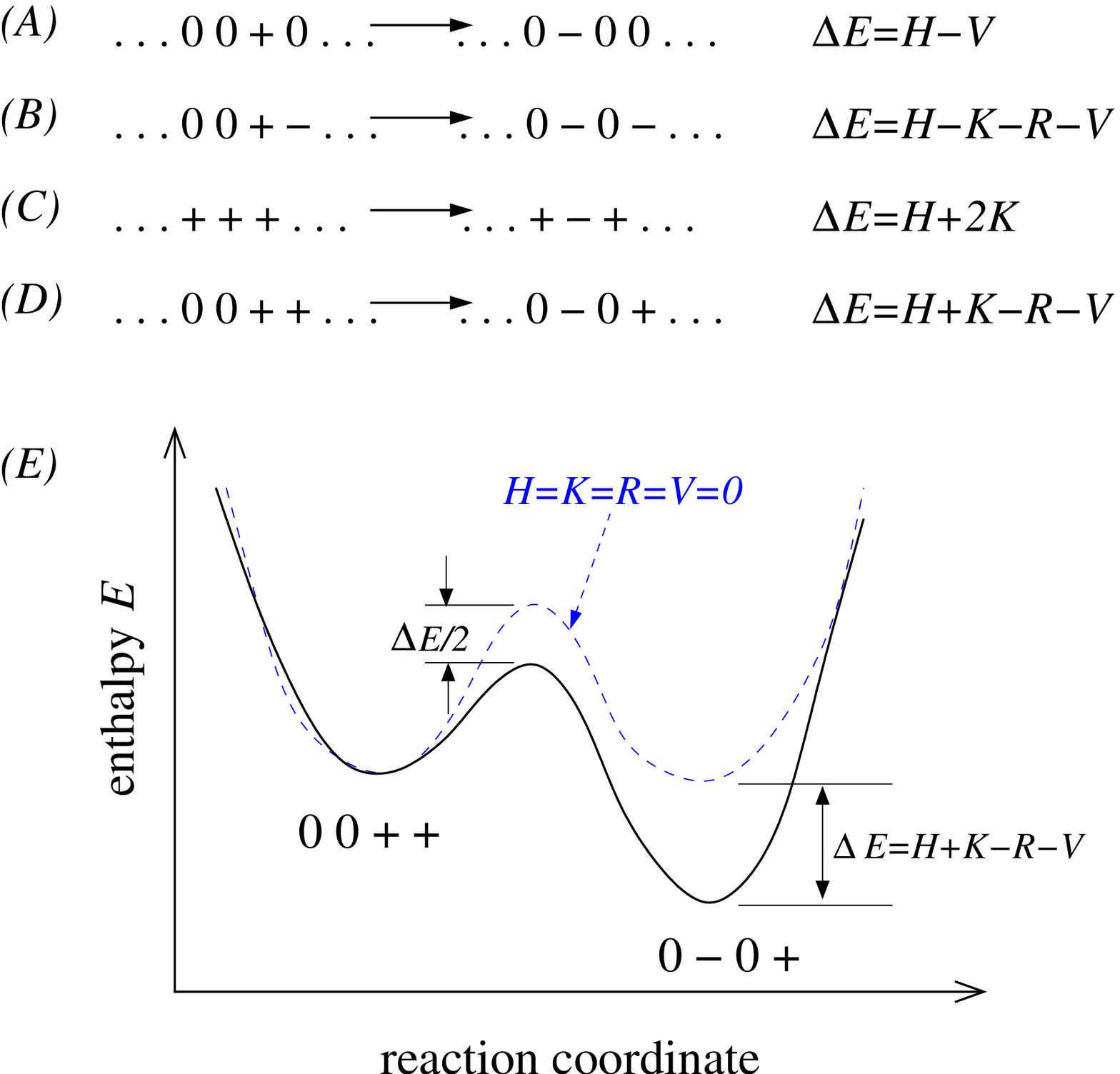}
\end{center}
\vspace{-2mm}
\caption{($A-D$) Energy differences between final and initial states 
which involve a change in ferroelectric coupling, net dipole 
moment, and repulsive interactions. ($E$) A representative energy 
barrier profile for $H=K=R=V=0$ (dashed curves). The energy profile for 
$H,K,R,V \neq 0$ for a transition between the states 
considered in ($D$) is shown by the thick solid curve.}
\label{deltaE}
\end{figure}

All eight ``parameters'' used in  our model (the rates $p_{\pm}, k_{\pm}, \a, \b, \g, \d$), can be related to
measured bulk quantities or derived from short-time MD simulations.  They are a minimal set and are
equivalent to the numerous bulk parameters used in other models \cite{SHU}, such as the bulk proton
diffusion constant, water orientational diffusion constants, etc.  Using similar MD approaches then,
one should be able to approximately fix the parameters used in our model. For example, variations in
the potential of mean force along  the pore (resulting from interactions of the different species with
the constituents of the pore interior) are embodied by site-dependent transition rates $p_{\pm}$ and
$k_{\pm}$.  Thus,  MD-derived potentials of mean force used in previous models can also be
implemented within our lattice framework. Such effects of local inhomogeneities in the hopping rates
have been studied analytically and with MC simulations in related models \cite{KOLODEFECT}.  

The basic model described above has been studied analytically in certain limits
where exact asymptotic results for the steady-state proton current $J$ were
derived \cite{CHOUJPA}.  However, this study did not
explicitly include any interactions other than proton exclusion and proton transfer
onto properly aligned water dipoles.  Effects arising from forces such as
repulsion between protons in close proximity, interactions between water dipoles
and external electric fields, and dipolar coupling between neighboring waters 
need to be considered.

In Fig \ref{deltaE}$A$, a proton moves down the electric
potential reducing the total enthalpy by $V$, and a
right-pointing dipole is converted into a left-pointing dipole at
an energy cost of $H$. Since both initial and final states have
adjacent, repelling protons, the repulsion energy $R$ does
not enter in the overall energy change. In Fig. \ref{deltaE}$B$,
a proton moves down the potential $(-V)$, a $``+''$ water is converted
to a $``-''$, $(+H)$, a dipole ``domain wall'' is removed $(-K)$,
and the repulsive energy between adjacent charged protons
is relieved $(-R)$. The representation of these nearest neighbor effects can be
succinctly written in terms of the energy of a specific
configuration

\begin{equation}
\begin{array}{ll}
E[\{\sigma_{i}\}] & \displaystyle = -K\sum_{i=1}^{N-1}\s_{i}\s_{i+1} - 
H\sum_{i=1}^{N}\s_{i} + \\[13pt]
\: & R\sum_{i=1}^{N-1}(1-\s_{i}^{2})(1-\s_{i+1}^{2})
-V\sum_{i=1}^{N}i(1-\s_{i}^{2}).
\label{HAMILTONIAN}
\end{array}
\end{equation}

\noindent The $H,K,R,V$ parameters used in $E[\{\sigma_{i}\}]$ are all in units of 
$k_{B}T$ and represent 

\begin{enumerate}
\item[$\bullet$] $H$: energy cost for orienting a water dipole against external field
\item[$\bullet$] $K$: energy cost for two oppositely oriented, adjacent dipoles
\item[$\bullet$] $R$: repulsive Coulombic energy of two adjacent protons
\item[$\bullet$] $V$: energy for moving a charged proton one lattice site against an external field.
\end{enumerate}

\noindent $V$ as the change in potential that a proton incurs as it moves between adjacent waters. 
The total transmembrane potential $V_{membrane} = NV$.  The local dielectric environment across
a channel can induce a spatially varying effective potential $V_{1\leq i \leq N}$
\cite{EDWARDS,Jordan84,Jordan92,INVALID}. In this study, we neglect this variation and assume
constant $V$ across the lattice.

In order to connect the quantities $H,K,R$,and $V$ to the rates $\a, \b, \g, \d, p_{\pm}, k_{\pm}$, we
will assume the transitions occur over thermal barriers.  Although barriers to proton hopping may be
small, we employ the Arrhenius forms in order to obtain a simple relationship so that qualitative
aspects of the effects of $H,K,R$, and $V$ can be illustrated. Activation-energy-based 
treatments for conduction across gramicidin channels have been 
previously studied \cite{CC02}. When the more complicated
interactions and external potentials are turned on, the effective transition rates $\xi \equiv \{\a,\b,\g,\d,
p_{\pm}, k_{\pm}\}$ on which we base our Metropolis Monte-Carlo become

\begin{equation}
\xi = \xi_{0} \exp\left({\Delta E \over 2}\right),
\label{XI}
\end{equation}

\noindent where $\xi_{0} \equiv \{\a_{0},\b_{0},\g_{0},\d_{0}, p_{0}, k_{0}\}$ are rate prefactors
when $H,K,R,V$ and $\Delta E$ are zero.  In defining Eq. \ref{XI}, we have assumed that the
energy barrier due to the difference $\Delta E=E[\{\s_{i}'\}]-E[\{\s_{i}\}]$ ($\{\s_{i}'\}$ and
$\{\s_{i}\}$ are the final and initial state configurations, respectively) is evenly split between
the barrier energies in the forward and backward directions. We use the convention that
$p_{+}=p_{-}=p_{0}$ and $k_{+}=k_{-}=k_{0}$ when $V=0$ and $H=0$, respectively. The
constraints and the state-dependent transition rates determined by Eqs.  1 and 2 completely
define a nonequilibrium dynamical model which we study using MC simulations.  Note that in
the original model (Fig. 2) we {\it do not} assume transition barriers, but rather only that the
dynamics are Markovian.

We first gained insight into the dynamics by considering
numerical solutions to the full Master equation for a short three site ($N=3$)
channel.  If we explicitly enumerated all $27 = 3^{3}$ states of the three site
model, the Master equation for the 27 component state vector $\vec{P}$ is 

\begin{equation}
{\dd \vec{P}(t) \over \dd t} = {\bf M}\vec{P}(t),
\end{equation}

\noindent where ${\bf M}$ is the transition matrix constructed from the rates 
$\xi$.  In steady-state, the $P_{i}$
are solved by inverting ${\bf M}$ with the constraint $\sum_{i=1}^{27}P_{i} =
1$. The steady-state currents are found from the appropriate elements in
$P_{i}$ times the proper rate constants in the model. For example,  if the
probability that the three-site chain is in the configuration $(+-0)$ is denoted
$P_{12}$, then the transition rate to state $P_{13} \equiv (+--)$ (corresponding
to the ejection of a proton from the last site into the right reservoir) is $\beta
e^{V-H-K}$ and the steady state current $J = \beta \sum_{i}'P_{i}$ (where 
the sum $\sum_{i}'$ runs over all configurations that contain a 
proton at the last site), will contain the term $\beta^{V-H-K}P_{12}$.

Monte-Carlo simulations were implemented for relatively small ($N = 10$) systems by
randomly choosing a site, and making an {\it allowed} transition with the probability $\xi
\exp(E_{i}-E_{f})/r_{max}$, where $r_{max}$ is the maximum possible transition rate of the
entire system.  In the next time step, a particle is again chosen at random and its possible
moves are evaluated. The currents were computed after the system reached steady-state
by counting the net transfer of protons across all interfaces (which separate adjacent sites
and the reservoirs) and dividing by $N+1$.  Physical values of $J$ are recovered by
multiplying by $r_{max}$. Particle occupation statistics within the chain were tracked
by using the definitions  of $+, 0$ and $-$ particle densities  
at each site $i$: $\rho_{+}(i)
= \langle \s_{i}(\s_{i}+1)/2\rangle, \rho_{0}(i)= \langle (1-\s_{i}^{2})\rangle$, and $\rho_{-}(i) =
\langle \s_{i}(\s_{i}-1)/2\rangle$, respectively.  However, for our subsequent discussion, it
will suffice to analyze simply the chain-averaged proton concentration $\bar{\sigma}_{0} =
\sum_{i=1}^{N} \rho_{0}(i)$. All MC results were checked and compared with the exact
numerical results from the three-site, 27-state master equation.  

\vspace{4mm}
\noindent {\bf RESULTS AND DISCUSSION}
\vspace{1mm}

Here, we present MC simulation results for a lattice of size $N=10$.  The mechanisms responsible for
the different qualitative behaviors are revealed and the effects of each interaction term will be
systematically analyzed. We explore  a range of relative kinetic rates, all nondimensionalized in units
of $p_{0}$, the intrinsic proton hopping rate from between adjacent waters.  Estimates for $p_{0}$
derived from quantum MD simulations are on the order of 1ps$^{-1}$
\cite{CHENG,Berendsen95,Klein98,SCHMITT}.

\begin{figure}
\begin{center}
\includegraphics[height=4.0in]{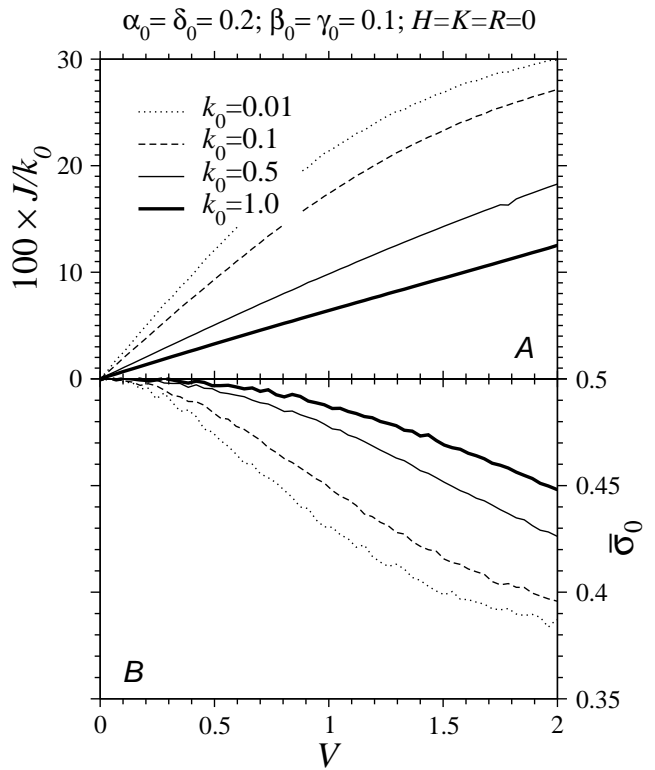}
\end{center}
\vspace{-2mm}
\caption{Saturation due to small flip rates $k_{+}=k_{-}=k_{0}$. 
Currents and rates  in all plots are nondimensionalized by units of 
$p_{0}$.
($A$) Small $k_{0}$ determines 
the rate limiting step whereupon increasing $V$ does little to 
increase the current. Increasing $k_{0}$ pushes the sublinear (saturation) regime 
of the $J-V$ relationship to larger 
values of voltage $V$. ($B$) The total proton occupancy decreases with decreasing 
$k_{0}$.}
\label{kV}
\end{figure}

One of the main features we wish to explore is the effect of multiple proton occupancy on
current-voltage relationships.  To understand what values of transition rates would permit multiple
proton occupancy, consider water at pH=7, which  has 10$^{-7}$M proton and hydroxyls.  This
concentration corresponds to about 60 H$_{3}$O$^{+}$ and 60 OH$^{-}$ species per cubic micron. 
Even at  pH 4, one would only have $\sim 60,000$ hydroniums per $\mu$m$^{3}$, corresponding to a
typical distance between hydroniums of $\sim 25$nm.  Since there are only $\sim 10-20$ waters
within a single-file channel, and at pH 4, only about one in 500,000 waters are protonated in bulk,
multiple protons in a single channel can occur only if protonated species within the channel are highly
stabilized by interactions with the chemical subgroups comprising the pore interior. This stabilizing
effect is modeled by small escape rates $\beta_{0}, \gamma_{0}$, and assumed to be distributed
equally such that $p_{0}$ remains constant across all sites within the lattice. Although from a
concentration point of view, small entrance rates $\a_{0}, \delta_{0}$ arise from infrequent protons
that wander into the first site of the channel, their exit rates $\b_{0}, \gamma_{0}$ can be suppressed
even more by their stabilization once inside the channel.  In all of our simulations, we will assume
proton stabilization is moderately strong and limit ourselves rates $\b_{0},\g_{0} < \a_{0},\d_{0}$. The
values we use give steady-state proton occupancies across the whole  range of values from
$\lesssim 1$ to $N$.  

%In bulk solution, excess protons are thought to protonate the lone pairs on water
%oxygens and may exist as small clusters ({\it e.g.} Eigen complexes).  

First consider symmetric solutions and featureless, uniform pores where $\a_{0}=\d_{0},
\b_{0}=\g_{0}$. The only possible driving force is an external voltage $V$.  In
Fig.  \ref{kV}, we plot the current-voltage relationship for various flipping rates
$k_{0}$. We initially ignore interaction effects and set $H=K=R=0$.  Currents
for sufficiently small $V$ are always nearly linear. However, for sufficiently
large $V$, the rate limiting step eventually becomes the water flipping rate
$k_{0}$. Further increases in $V$ do not increase the overall steady-state current,
and the current-voltage curve becomes sublinear before saturating.  The
crossover to sublinear (water flipping rate limited) behavior depends on the
value of $k_{0}$, with sublinear onset occurring at higher voltages $V$ for
larger $k_{0}$.  In the noninteracting case, for most reasonable values of rate
constants, any possible superlinear regime does not arise as it is washed out
by the sublinear, water flip rate-limited saturation.  The only instance found
where noticeable superlinear behavior in the steady-state proton current arises
is in the limit of large $k_{0}$ {\it and} when $\a_{0},\d_{0}, p_{0} \ll \beta_{0},
\gamma_{0}$.  For the parameters explored, the currents $J$ increase with
increasing $k_{0}$ (Fig.  \ref{kV}$A$); thus, the mean proton occupancies
are {\it qualitatively} consistent with dynamics limited by internal proton hops.
For small flipping rates, successive entry of protons is slow, while exit is not
affected. As $k_{0}$ is increased, the bottlenecks near the entrance are relieved
to a greater degree than those near the exit, increasing the overall proton
occupancy  (cf. Fig.  \ref{kV}$B$).

\begin{figure}
\begin{center}
\includegraphics[height=4.0in]{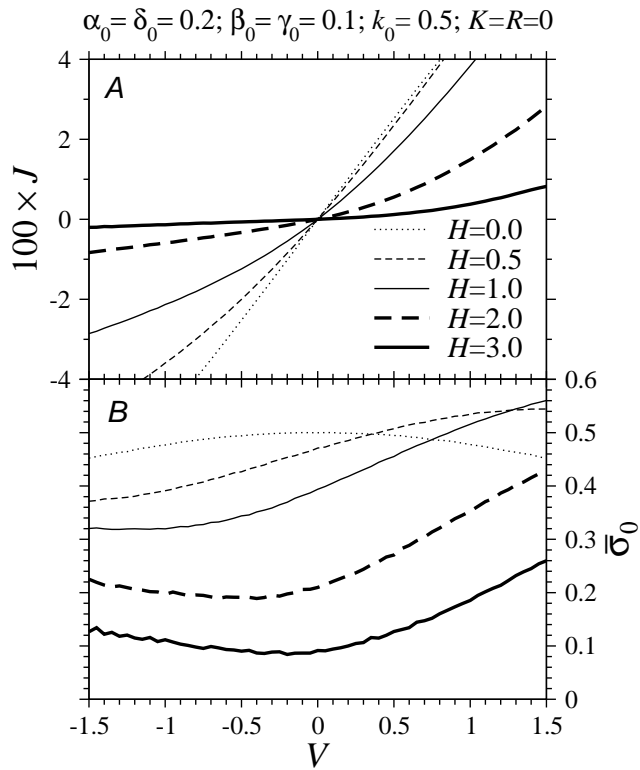}
\end{center}
\vspace{-2mm}
\caption{Currents ($A$) and averaged proton occupation ($B$) in the
presence of a constant water dipole-aligning field $H>0$. For larger $V$, 
the $V$-independent $H$ assumption used in this scenario
will break down due to the orientation effects of $V$ on the water dipoles.}
\label{HFIXED}
\end{figure}

Figure \ref{HFIXED} displays the effects of a fixed, external, dipole-orienting field
$H\neq 0$. All other interactions and fields, except the external driving voltage $V$,
are turned off. The convention used in the energy Eq. \ref{HAMILTONIAN} favors a
``+'' state for $H>0$. This asymmetry leads to an asymmetry in the $J-V$ relationship
(Fig. \ref{HFIXED}$A$).  After an initial proton has traversed the channel, flipping of
the  ``$-$'' waters left in its wake is suppressed for $H>0$, thereby preventing further net
proton movement.  The persistent  blockade induced by increasing $H$ is evident in
Fig. \ref{HFIXED}$B$ where the proton density decreases for increasing $H$.

\begin{figure}
\begin{center}
\includegraphics[height=4.0in]{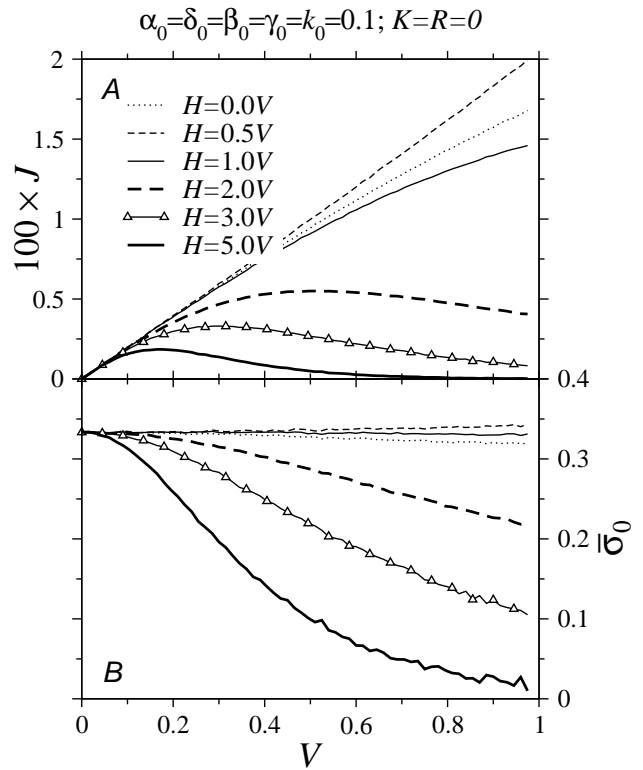}
\end{center}
\vspace{-2mm}
\caption{($A$) Negative differential resistance (NDR) for large $L_{HV}, V$. 
Although transitions such as $\ldots -+0-+\ldots \rightarrow \ldots -+0++\ldots$
are accelerated, giving rise to a state where proton transport to the right 
is possible, NDR can arise because transitions such as 
$\ldots -+0++\ldots \rightarrow  \ldots -+-0+\ldots$ created an additional 
$-$ particle and is disfavored. ($B$) The average proton occupation {\it decreases} 
as $V$ for large $L_{HV}$.}
\label{HV}
\end{figure}

Although $H$ is held fixed in Fig. \ref{HFIXED}, physically, dipole alignment fields arise from
external electric fields that couple to the permanent dipoles of water.  Therefore, we expect
that $H = L_{HV}V$ where $L_{HV}$ represents the orientational polarizability of the water
molecule.  It has been conjectured that  when $L_{HV}$  is positive (defined as preferring waters
with lone pairs pointing to the left, or in the ``$+$'' state), the current should increase
superlinearly with $V$ since waters ahead of any proton will be oriented properly as to receive
it.  Figure \ref{HV} shows the current-voltage relationship for various $L_{HV}$.  Although for
very small $L_{HV}$, the current does increase very slightly, it becomes severely sublinear for
larger  $L_{HV}$ and $V$. In fact, it can attain a negative differential resistance (NDR) similar to
that found in Gunn diodes or other ``negistor'' devices. The physical origins of NDR in proton
conduction arise from the energetic cost of producing a ``$-$'' state as a proton moves forward.
Although the path ahead of the proton is biased to ``$+$'' states, the proton transfer step as
defined in our model {\it necessarily} leaves behind a ``$-$'' particle. Thus, although the field
$H=L_{HV}V$ properly aligns waters ahead of a proton, it also provides an energy cost for the
tail of ``$-$'' particles left by a forward-moving proton.  This energetic penalty inhibits the proton
from moving forward despite the direct driving force $V$ acting on it.

The average density plotted in Fig. \ref{HV}$B$ decreases as $V$ or large $L_{HV}$. Large
$L_{HV}$ not only hinders forward proton hops, but enhances backward hops of protons that
have just hopped forward during its previous time step. Proton dynamics are slowed
dramatically, and only at the last site can they exit the pore. Proton entry from the left reservoir
on the other hand, is often quickly followed by exit back into the left reservoir. The protons are
{\it effectively} entry-limited, and the density is rather low. As $V$ increases, the dynamics
become even more ``entry-limited,'' and the overall proton occupancy decreases.

\begin{figure}
\begin{center}
\includegraphics[height=4.0in]{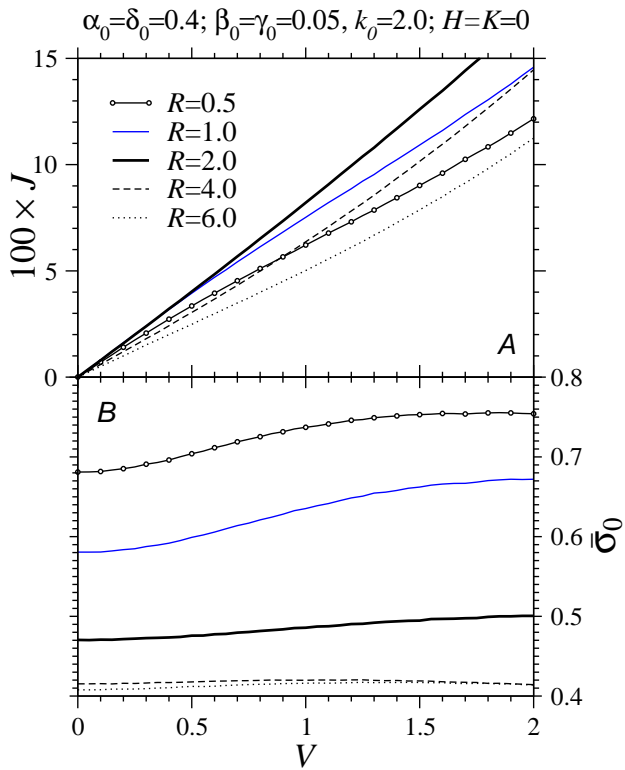}
\end{center}
\vspace{-2mm}
\caption{The effects of increasing nearest-neighbor proton-proton
repulsion within the chain. Fixed parameters are 
$\a_{0}=\d_{0}=0.4$, $\b_{0}=\g_{0}=0.05$, $k_{0}=2.0$, and 
$H=K=0$. ($A$) The onset of sublinear behavior
in the $J-V$ relationship is delayed for larger 
repulsions $R$, making the curves appear locally more superlinear.
($B$) The average proton densities per site. For small $R$, 
although densities are high, increasing $V$ increases the 
clearance rate near the entrance such that the 
effectively increased injection increases overall 
proton density. At higher repulsions $R$, the 
clearance effects is not as strong and the
simultaneously increased extraction rate 
prevents a large increase in the overall proton density.}
\label{ARV}
\end{figure}

\begin{figure}
\begin{center}
\includegraphics[height=4.0in]{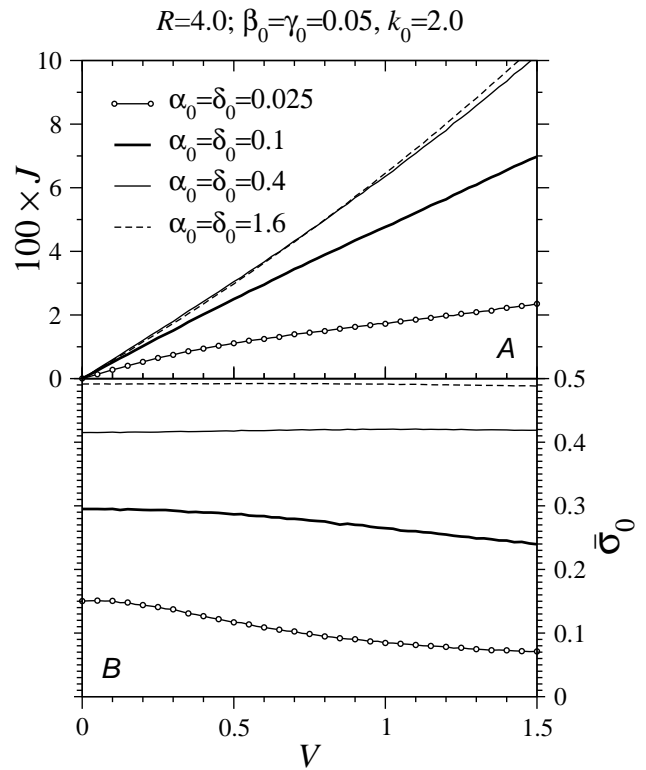}
\end{center}
\vspace{-2mm}
\caption{Transition from sublinear to superlinear current behavior as 
proton concentration in the symmetric reservoirs is increased.
($A$) $J-V$ relationship for various concentrations $\a_{0}=\d_{0}$ for 
fixed $H=K=0,R=4.0$, $\b_{0}=\g_{0}=0.05$, and $k_{0}=2.0$. ($B$) The averaged proton 
concentration $\s_{0}$ at each lattice site as a function of driving voltage.
The concentrations increase for all ranges of $V$ as $\a=\d$ is increased.}
\label{RAV}
\end{figure}

The effects of proton-proton repulsion ($R>0$) are considered in Figs \ref{ARV} and \ref{RAV}.
These simulations are consistent with the hypothesis that proton-proton repulsions can give
rise to superlinear current \cite{HIL78}. Figure \ref{ARV}$A$ shows a slight preference for
superlinear behavior as repulsion $R$ is increased. Not surprisingly, Fig.  \ref{ARV}$B$ shows
that the overall density of protons within the pore decreases with increasing repulsion.  

The sublinear-to-superlinear behavior as the proton concentration in the identical reservoirs
is increased is shown in Fig. \ref{RAV}$A$.  Although for these parameters, the effect is not
striking, there is indeed a trend away from sublinear behavior as pH is decreased, or, as
$\a_{0}=\d_{0}$ is increased. Measurements, though, also show rather modest superlinear
behavior \cite{EISENMAN80,PHILLIPS,ROKIT}. The occupancy also increases with
decreasing pH, enhancing the effect of proton-proton repulsion.  These behaviors are
consistent with experimental findings \cite{EISENMAN80} and those in the simulations
depicted in Fig.  \ref{ARV} where increased repulsion exhibited superlinear $J-V$ curves.

\begin{figure}
\begin{center}
\includegraphics[height=4.0in]{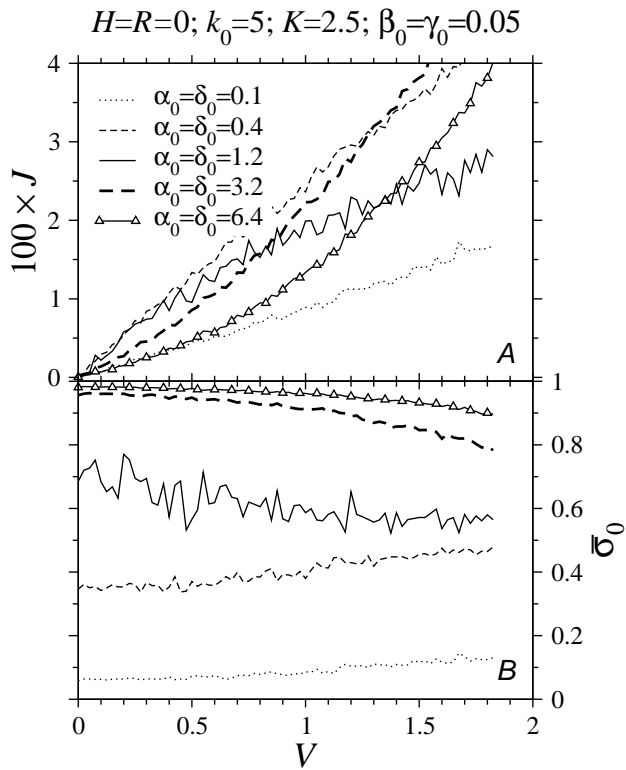}
\end{center}
\vspace{-2mm}
\caption{($A$) The current-voltage relationship for various
proton injection rates in the presence of 
ferromagnetic water dipole coupling. ($B$) Mean proton occupations increase with 
increasing injection rates.}
\label{FIGK}
\end{figure}

Finally, we consider the effects of dipole coupling $K\neq 0$ between adjacent water
molecules.  This interaction is analogous to a nearest neighbor ferromagnetic coupling in {\it
e.g.}, Ising models.  Fig. \ref{FIGK}$A$ shows that for sufficiently large $\a_{0}=\d_{0}$, a
superlinear behavior arises (for small enough $V$ and large enough $k_{0}$ such that saturation
has not yet occurred).  Notice that as $\a_{0}=\d_{0}$ is increased, the $J-V$ relationship can
become more sublinear before turning superlinear. Here, we have used a higher value of $k_{0}$
to suppress sublinear behavior to larger $V$, but the qualitative shift from sublinear to slightly
superlinear behavior exists for small $k_{0}$.  Moreover, recent comparisons between
gramicidin A and gramicidin M channels suggest that water reorientation is not rate-limiting
\cite{GOWEN}.  The nature of the superlinear behavior can be deduced from Fig. 
\ref{FIGK}$B$, where the mean proton density is shown to increase with $\a_{0}=\d_{0}$.
Waters that neighbor a proton are relieved of their dipolar coupling and can more readily flip to
a configuration that would allow acceptance of another proton. For example, the transition
$\ldots 0 - 0 \ldots \rightarrow \ldots 0 + 0 \ldots$ will occur faster than $\ldots - - 0 \ldots
\rightarrow \ldots - + 0 \ldots$. This lubrication effect arises only when the proton density is
high and  $K\neq 0$.

\vspace{4mm}
\noindent {\bf SUMMARY AND CONCLUSIONS}
\vspace{1mm}

We have developed a lattice model for proton conduction that quantifies the kinetics among
three approximate states of the individual water molecules inside a simple, single-file channel
such as gramicidin A. The three states represent water molecules with left and right-pointing
water dipoles, and protonated ions.  Our approach allows us to explore the {\it steady-state}
behavior of proton currents, occurring over timescales  inaccessible by MD simulations. The
model, along with analyses of Monte-Carlo simulations, also extends  analytic models
\cite{Schumaker99,SHU} to include multiple proton occupancy and the memory effects of
protons that have recently traversed the water-wire.  Monte-Carlo simulations of the lattice
model was performed to test conjectures on a number of observed  qualitative features in
proton transport across water wires. Four interaction energies that modify the kinetic rates
are considered: A dipole-orienting field which tends to align the water molecules, a
ferromagnetic dipole-dipole interaction terms between neighboring water molecules, a
penalty from the repulsion between neighboring protons, and a external electric field
(transmembrane potential) that biases the hops of the charged protons.

We find current-voltage relationships that can be both superlinear and sublinear
depending on the voltage $V$. For large enough voltages, the proton hopping step
is no longer rate limiting. Water flipping rates limit proton transfer and
further increases in $V$ do little to increase the steady-state proton current
$J$. This observation suggests that the observed transition from sublinear to
superlinear behavior can be effected by varying an {\it effective} water flipping
rate.  Although we find that indeed proton-proton repulsion can lead to slightly
superlinear $J-V$ characteristics, particularly for large repulsions and proton
injection rates (low pH). 

Dipole-dipole interactions between neighboring waters are also incorporated. 
Previous single-proton theories \cite{Schumaker99,SHU} have
considered the propagation of a single defect back and forth in the pore.  In our
model, the number of protons and defects are dynamical variables that depend on
the injection rates and the dipole-dipole coupling, respectively.  For large coupling
$K$, we expect very few defects, and effective water flipping rates will be low. 
However, when injection rates and proton occupancy in the pore is high, some
dipole-dipole couplings are broken up by the intervening protons.  Thus, protons
can ``lubricate'' their neighboring dipoles, allowing them to flip faster than if they
were neighboring a dipole pointed in the same direction. Using simulations, we
showed that this lubrication effect can  give rise to a superlinear $J-V$ relationship 

Although the parameters used in our analyses can be further refined by estimating them from shorter
time MD simulations, or other continuum approaches \cite{EDWARDS,Jordan92}. More complicated
local interactions with membrane lipid dipoles \cite{ROKIT} and internal pore constituents (such as
Trp side groups \cite{BUSATH99,GOWEN}) can be incorporated by allowing $H, K, p_{0}$ and/or
$k_{0}$ to reflect the local molecular environment by varying along the lattice site (position) within the
channel \cite{KOLODEFECT}.

\vspace{4mm}

The author thanks Mark Schumaker for vital discussions and comments on
the manuscript. This work was performed with the support of the National Science
Foundation through grant DMS-0206733, and the National Institutes of Health through grant
R01 AI41935.

\vspace{4mm}

\begin{appendix}

%\noindent {\bf APPENDIX A: NOINTERACTING MEAN-FIELD RESULTS}

\section{NOINTERACTING MEAN-FIELD RESULTS} 

For the sake of completeness, and as a guide to aid qualitative understanding, we review analytic
results in the case $R=K=H=0$, where only exclusions are included.  Some of these results have been
derived previously using mean-field approximations \cite{CHOUJPA}.  

If $V=0$ ($\xi = \xi_{0}$), only pH differences between the two reservoirs can affect a nonzero
steady-state proton current. The proton concentration difference is reflected by a difference
between the entry rates from the two reservoirs $\a_{0}\neq \delta_{0}$, and the steady-state
current can be expanded in powers of $1/N$: $J = a_{1}/N + a_{2}/N^{2} +
{\cal O}\left(N^{-3}\right)$.  In the long chain limit, we found  \cite{CHOUJPA}

\begin{equation}
\begin{array}{l}
J \sim {k_{+}k_{-}\over N(k_{+}+k_{-})}\times \\[13pt]
\ln \left[{\b(k_{+}+k_{-})+k_{+}\d\over \g(k_{-}+k_{+})+k_{-}\a}{\g(k_{+}+k_{-})+\a(p_{-}+k_{-})\over
\b(k_{+}+k_{-})+k_{+}\d(p_{-}/k_{-}+1)}\right] + {\cal O}(N^{-2}).
\end{array}
\label{J2}
\end{equation}

\noindent For channels with reflection-symmetric molecular structures, 
$\b_{0}=\g_{0}$, and Eq. \ref{J2} can be further simplified by expanding in powers of 
$k_{-}\a - k_{+}\d$,

\begin{equation}
\begin{array}{l}
\displaystyle  J \sim {\b p_{+}k_{-}(k_{-}\a - k_{+}\d) \over 
N \left[\b(k_{-}+k_{+})+k_{+}\d\right](\b+\d) (k_{+}+k_{-})} +\\[13pt]
\:\hspace{2cm} \displaystyle  O\left((k_{-}\a-k_{+}\d)^{2}\right)
+{\cal O}(1/N^{2}),
\end{array}
\label{J2b}
\end{equation}

\noindent Finally, in the {\it large} $\a$ and $\d = 0$ limit, 

\begin{equation}
\begin{array}{l}
\displaystyle J \sim {k_{+}k_{-} \over (k_{+}+k_{-})N} \log \left(1+{p_{-}\over k_{+}}\right) 
- \\[13pt]
\: \hspace{2cm}\displaystyle  {\gamma k_{+}k_{-}p_{-} \over 
\a N (k_{+}+k_{-})(k_{+}+p_{-})} + {\cal O}(\a^{-2}N^{-1}).
\end{array}
\end{equation}

For driven systems, where, say, $\a > \d, \b>\g$, and $p_{+}> p_{-}$, a finite current persists
in the $N\rightarrow \infty$ limit. We can use
mean-field approximations familiar in the totally asymmetric simple exclusion process (TASEP)
\cite{DER98,SCHUTZ} to conjecture that three current regimes exist.  If the both proton
entry and exit is fast, and the rate-limiting steps involve water flipping, or interior protons
hops with rate $p_{+}$, we expect that a maximal current regime exists and that the densities of
the three states along the interior of a long chain are spatially uniform.
Mean-field analysis from previous work \cite{CHOUJPA} yields

\begin{equation}
\begin{array}{l}
\displaystyle  J = {2(p_{+}k_{-}-p_{-}k_{+}) \over (p_{+}+p_{-})^{2}}
\bigg[{(p_{-}+p_{+})\over 2}+k_{-}+k_{+} - \\[13pt]
\: \hspace{2cm} \displaystyle  \sqrt{k_{+}+
k_{-}}\sqrt{k_{-}+k_{+}+p_{+}+p_{-}}\bigg].
\label{JM}
\end{array}
\end{equation}

\noindent For a purely asymmetric process, $p_{-}=0$, and 
the current approaches the analogous maximal-current 
expression of the single species TASEP,

\begin{equation}
J(p_{-}=0) \sim {p_{+}k_{-} \over 4(k_{-}+k_{+})} + {\cal O}\left({p_{+}\over k_{-}}\right),
\end{equation}

\noindent except for the additional factor of $k_{-}/(k_{-}+k_{+})$ representing
the approximate  fraction of time sites ahead of a proton are in the $+$
configuration. These approximations neglect the influence of protons that have
recently passed, temporarily biasing the water to be in a ``$-$'' configuration.
Therefore, it is not surprising that these results are accurate only in the $k_{+},
k_{-} \gg p$ limit.

A similar approach is taken when the currents are entry or exit limited.
From the mean-field approximation of the steady-state equation for
$\rho_{\pm}$ near the channel entry,

\begin{equation}
\begin{array}{l}
\displaystyle  {\partial \rho_{-} \over \partial t} = p_{+} \rho_{0}\rho_{+} + k_{+}\rho_{+}-
k_{-}\rho_{-} = 0\\[13pt]
\displaystyle  {\partial \rho_{+} \over \partial t} = -\a \rho_{+}-k_{+}\rho_{+}+k_{-}\rho_{-}=0,
\label{RLLEFT}
\end{array}
\end{equation}

\noindent where we have for simplicity set $p_{-}=\gamma=0$. Upon using
normalization $\rho_{-}+\rho_{0}+\rho_{+} = 1$, and Eqs. \ref{RLLEFT}, we find the 
mean densities near the left boundary

\begin{equation}
\begin{array}{l}
\displaystyle \rho_{-} = {(\a+k_{-})(p_{+}-\a) \over p_{+}(\a+k_{-}+k_{+})}, \\[13pt]
\displaystyle \rho_{+} = {k_{-}(p-\a) \over p_{+}(\a+k_{-}+k_{+})},
\end{array}
\end{equation}

\noindent and the approximate entry rate-limited steady-state current

\begin{equation}
J\approx p_{+}\rho_{0}\rho_{+} = \a\rho_{+} = 
{\a k_{-}(1-\a/p)\over (\a+k_{-}+k_{+})}.
\end{equation}

\noindent This result resembles the steady-state current of the low density phase in the simple
exclusion process \cite{DER98,MRNA2}, except for the factor $k_{-}/(\a+k_{-}+k_{+})$ representing
the fraction of time the first site is in the $+$ state, and able to accept a proton from the left reservoir.

When the rate $\beta$ is rate-limiting, we consider the mean-field equations 
near the exit of the channel

\begin{equation}
\begin{array}{l}
\displaystyle  {\partial \rho_{-} \over \partial t} = \b \rho_{0}
+k_{+}\rho_{+}-k_{-}\rho_{-}=0 \\[13pt]
\displaystyle  {\partial \rho_{+} \over \partial t} = 
-p_{+}\rho_{0}\rho_{+}+k_{-}\rho_{-}-k_{+}\rho_{+}=0,
\label{RLRIGHT}
\end{array}
\end{equation}

\noindent and their solutions 

\begin{equation}
\rho_{-} = {\b (k_{+}+p_{+}-\b)\over p_{+}(k_{-}+\b)}, 
\quad \rho_{+}={\b \over p_{+}}.
\end{equation}

\noindent The exit-limited steady-state current is thus

\begin{equation}
J \approx \b\rho_{0} = {\b \over k_{-}+\b}\left(k_{-}
-{\b (k_{-}+k_{+})\over p_{+}}\right).
\end{equation}

The results above are derived from mean-field assumptions which neglect correlations in particle
occupancy between neighboring sites. Although mean-field theory happens to give exact results for
the simple exclusion process, the results above are only exact in the large $k_{\pm}/p_{\pm}$ limit, as
has been shown by Monte-Carlo simulations \cite{CHOUJPA}.  Only in this limit, where the memory
of a previously passing proton is quickly erased, are the mean-field results quantitatively accurate
\cite{CHOUJPA}.  Nonetheless, the mean-field calculations of the simplified system $(H=K=R=0)$
yields qualitatively correct results for the steady-state current, provides a connection with
well-known results of the TASEP, and gives an explicit qualitative description of the mechanisms at
play.

\end{appendix}

\end{document}